# *Laser induced nuclear waste transmutation*

Charles Hirlimann

IPCMS Unistra-CNRS UMR 75004, 23 rue du Lœss, BP 43
F-67034 Strasbourg cedex2

This document is the proposal for a study that was supported in 2002 by University Louis Pasteur in Strasbourg, France, exploring the feasibility of a new way for taking care of nuclear waste using lasers. This work has been performed by a multidisciplinary r-team: Marie-Paule Baumann and Gérard Rudolf, nuclear physicists, Isabelle Billard and Klaus Lutzenkirchen, nuclear chemists, and Olivier Crégut and I, laser physicists. I have been leading this team from 2000 to 2003. This text appears in: "Lasers et Technologies Femtosecondes", M. Sentis and O. Utéza Ed., (Publications de l'Université de Saint-Etienne), p. 69-80 (2005).

## *Nuclear Waste*

When producing electricity that collects the mass energy that is available at the time of the induced disintegration of radioactive elements, other unstable elements are produced with half-life[1] span durations ranging from less than one second to hundreds of thousands of years and which are considered as waste. Managing nuclear waste with a half-life of less than 30 years is an easy task, as our societies clearly know how to keep buildings safe for more than a century[2], the time it takes for the activity to be divided by a factor of 8. High-activity, long-lasting waste that can last for thousands of years or even longer, up to geological time laps, cannot be taken care of for such long durations. Therefore, these types of waste are socially unacceptable; nobody wants to leave a polluted planet to descendants.

The French law (91-1391, 30 December 1991) has assigned to research agencies and nuclear economic actors the mandatory task of searching for practical ways to get rid of high-level activity and long-lasting nuclear waste. Three possible ways have been recognised: separation-transmutation of long-lasting elements, deep geological storage, and surface long-lasting storage[3]. The first line this study is aimed at establishing the best conditions for performing laser-assisted transmutation of long-lasting elements.

---

[1] The half-life of a radioactive element is the time that is needed for half a population of atoms to disintegrate.
[2] Some cathedrals are at least 500 hundred years old and have been taken care of for that long!

French nuclear plants are Pressured Water Reactors (PWR) producing radiotoxic elements as a side product of electricity. These elements include actinides and fission products. Overall, 90% of the actinides are made out of plutonium, the remaining part being 10% of minor actinides. Wastes don't have any precise absolute definition: a waste is a side product that cannot be used for any pertinent purpose. At some point a waste can be considered being a source of raw material! This is why, in France, plutonium is no waste as its oxide is, mixed with uranium oxide, forms the MOX (Mixed OXydes) nuclear fuel that is burnt in the PWRs. As Rapid Neutron Reactors (RNR) based on the disintegration of plutonium are envisioned for being the fourth generation of reactor it seems wise to save this metal. Each year, 10 tonnes of plutonium are produced in France. Among the minor actinides, americium and neptunium are strong producers of neutrons and gamma rays.

The second family of nuclear waste is made of the products resulting from the atomic fission process; namely technecium-99, iodin-129, and cesium-135. These elements have a long lasting half-life and even if they only are the $10^{-5}$ part of the exhausted nuclear fuel they largely contribute to the radio-toxicity of the nuclear waste, as they are highly soluble in water and highly mobile in geological layers.

Transmutation of nuclear wastes aims at changing instable elements into at least short lived ones. The way to transmutation is to irradiate the nuclei with elementary particles: protons or neutrons or with gamma rays. Gamma ray irradiation has rapidly been abandoned. Some technologic locks remained unsolved for producing intense gamma ray beams. Most research on transmutation explores neutron and proton irradiation; one of the main idea being taking advantage of the production of thermal protons in nuclear plants or high-energy neutron in the next generation of reactors. The following table shows the production of minor actinides in the French nuclear plants.

---

[3] Les déchets nucléaires, dossier scientifique de la Société Française de Physique sous la direction de RenéTurlay, EDP Sciences 1997.

| Actinide | Production | Half-life |
|---|---|---|
| Plutonium | 11 t/yr | — |
| Neptunium | 800 kg/yr | $^{237}$Np : 2.1 10$^6$ years |
| Americium | 250 kg/yr | $^{243}$Am : 7380 years |
| Curium | 0.5 kg/yr | $^{245}$Cm : 8500 years |

Even though the amount of fission products that are produced in nuclear plants is quite small, they can be deleterious to public health.

| Element | Production, kg/year | Half-life, years |
|---|---|---|
| Caesium | 4.8 | $^{135}$Cs, 2 10$^6$ |
| Iodine | 1.8 | $^{129}$I, 1.57 10$^7$ |

### *Gamma rays and powerful lasers*

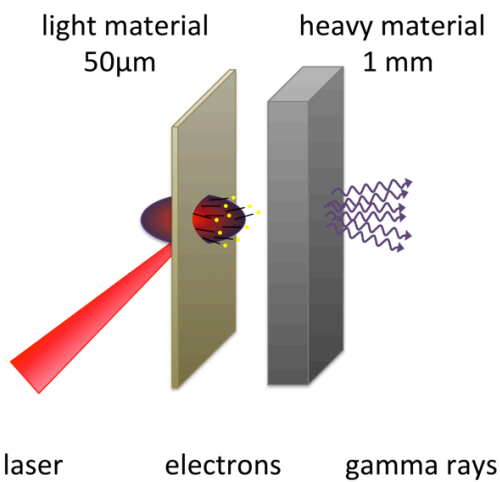

**Figure 1.** A laser beam is focussed down onto a thin foil of light atoms materials, creating a plasma in which electrons are accelerated to high energies. These electrons when stopped in a high atomic weight material do generate gamma rays.

At the very end of the last century, two teams, one British, headed by Ledingham[4], and the other one from the U.S., led by Cowan[5], published papers demonstrating the possibility of inducing nuclear reactions using short laser light pulses in the visible range. Laser-induced fission of actinides was demonstrated at the very beginning of the present century by a German team headed by Schwoerer[6].

The electric field which is created in the focal volume of an intense, 10$^{20}$ W/cm$^2$, focussed laser light pulse reaches a value as high as 10$^{11}$ V/cm. This value is much larger than the electric field to which the electron on the first Bohr state of a hydrogen atom is submitted due to the proton, which is "only" equal to 10$^5$ V/cm. The magnetic field of the electromagnetic wave is not any more negligible as it reaches a value of the order of $10^5$ Tesla. When such an intense light pulse impinges onto some piece of matter, it very rapidly creates plasma. Inside the plasma, the heavy positive ions can be considered as still (adiabatic approximation) while the electrons are accelerated by the electric field. The force applied to the electrons during ¼ of the cycle of the field is of the order of $10^{-6}$ Newton which corresponds to an acceleration of $1.6 \; 10^{24}$ m/s². This colossal acceleration pushes the electrons to relativistic velocities close to the speed of light in a time lap that is less than one femtosecond. As the magnetic field cannot be neglected, the electrons are submitted to a ponderomotive force that is the relativistic equivalent of a Lorentz force. This process moves the electrons very rapidly in the forward direction along the light propagation direction.

---

[4] K. W. D. Ledingham et al. Phys. Rev. Lett., **899** (2000).
[5] T. E. Cowan et al. Phys. Rev. Lett., **903** (2000). 4 H.
[6] H. Schwoerer et al. Europhys. Lett. **61** (2003).

The energy distribution of these accelerated electrons peaks a bit below an energy of 10 MeV and extends up to 00 MeV. When entering matter, the electrons are abruptly slowed down, and this slowing down is larger in atoms with a high atomic weight like tantalum or lead. This brutal deceleration is responsible for the emission (Brehmsstrahlung) of high-energy electromagnetic rays that are quasi-collimated in the forward direction[7]. These resulting gamma rays can be used to induce fission in unstable nuclei. Even if their photon energy is less than the potentials barrier heights for beta or alpha decay of radioactive products that are generally less than 10 MeV, they have recently been used for transmuting $^{129}$I (half-life = 15.7 million years) into $^{128}$I (half-life = 25')[8]. This clearly paves the way to laser-assisted nuclear waste transmutation.

The following table compares Nd doped lasers with titanium-sapphire lasers through the number of fissions/second that have been induced during published experiments.

| Gain medium | Energy J | Duration fs | Peak power TW | Intensity W/cm² | # shots s⁻¹ | # fissions s⁻¹ |
|---|---|---|---|---|---|---|
| Nd:glass Vulcan | 75 | 1000 | 100 | 1019 | 2 | 103 |
| Ti:Sap Jena, LOA | 0.5 | 80 | 15 | 1020 | 10 | 104 |

---

[7] P.A. Norreys et al. Physics of Plasmas **6**, 2150 (1999).
[8] K. W. D. Ledingham et al. J. Phys. D : Appl. Phys. **36**, L79 (2003).

As can be seen, the short duration laser is more efficient by one order of magnitude.

One light pulse lasting for 160 femtoseconds, carrying an energy of 200 mJ, and focussed down to $10^{18}$ W/cm$^2$ onto a $C_2D_4$ foil produces $10^{10}$ accelerated electrons, where energy spans from 1 to 10 MeV. When stopped in bulk aluminium, the $5 \cdot 10^7$ emitted per pulse gamma photons do span the same range of energies. Such a realistic gamma ray source has been used for producing neutrons even though its power is two orders of magnitude smaller than the power of the sources used for proving laser-induced fission[9].

**Note added at the time of the translation** (March 2016)**:** The general idea sustaining this work is that it would be safer to keep the nuclear waste confided inside the nuclear plants where it is generated. In this way, nuclear plants should add two more facilities: one for isotope separation and a large laser one.

---

[9] G. Pretzler et al., Phys. Rev. E, **58**, 1165 (1998).